\documentclass[aps,prd,twocolumn,10pt
,
tightenlines,
nofootinbib,showpacs]{revtex4-1}
\usepackage{amssymb,latexsym}
\usepackage{amsmath,amsbsy,bbm}
\usepackage{epsfig,bm}
\usepackage{graphicx,comment}
\DeclareMathOperator{\tr}{tr}

\begin{document}
\def\a{{\alpha}}
\def\b{{\beta}}
\def\d{{\delta}}
\def\D{{\Delta}}
\def\X{{\Xi}}
\def\e{{\varepsilon}}
\def\g{{\gamma}}
\def\G{{\Gamma}}
\def\k{{\kappa}}
\def\l{{\lambda}}
\def\L{{\Lambda}}
\def\m{{\mu}}
\def\n{{\nu}}
\def\o{{\omega}}
\def\O{{\Omega}}
\def\S{{\Sigma}}
\def\s{{\sigma}}
\def\th{{\theta}}

\def\ol#1{{\overline{#1}}}

\def\Dslash{D\hskip-0.65em /}
\def\Dtslash{\tilde{D} \hskip-0.65em /}

\def\CPT{{$\chi$PT}}
\def\QCPT{{Q$\chi$PT}}
\def\PQCPT{{PQ$\chi$PT}}
\def\tr{\text{tr}}
\def\str{\text{str}}
\def\diag{\text{diag}}
\def\order{{\mathcal O}}

\def\cF{{\mathcal F}}
\def\cS{{\mathcal S}}
\def\cC{{\mathcal C}}
\def\cB{{\mathcal B}}
\def\cT{{\mathcal T}}
\def\cQ{{\mathcal Q}}
\def\cL{{\mathcal L}}
\def\cO{{\mathcal O}}
\def\cA{{\mathcal A}}
\def\cQ{{\mathcal Q}}
\def\cR{{\mathcal R}}
\def\cH{{\mathcal H}}
\def\cW{{\mathcal W}}
\def\cM{{\mathcal M}}
\def\cD{{\mathcal D}}
\def\cN{{\mathcal N}}
\def\cP{{\mathcal P}}
\def\cK{{\mathcal K}}
\def\Qt{{\tilde{Q}}}
\def\Dt{{\tilde{D}}}
\def\St{{\tilde{\Sigma}}}
\def\cBt{{\tilde{\mathcal{B}}}}
\def\cDt{{\tilde{\mathcal{D}}}}
\def\cTt{{\tilde{\mathcal{T}}}}
\def\cMt{{\tilde{\mathcal{M}}}}
\def\At{{\tilde{A}}}
\def\cNt{{\tilde{\mathcal{N}}}}
\def\cOt{{\tilde{\mathcal{O}}}}
\def\cPt{{\tilde{\mathcal{P}}}}
\def\cI{{\mathcal{I}}}
\def\cJ{{\mathcal{J}}}

\def\eqref#1{{(\ref{#1})}}

\title{Isotensor Hadronic Parity Violation}
\author{B.~C.~Tiburzi}
\email[]{btiburzi@ccny.cuny.edu}
\affiliation{ Department of Physics,
        The City College of New York,  
        New York, NY 10031, USA \\
Graduate School and University Center,
        The City University of New York,
        New York, NY 10016, USA \\
RIKEN BNL Research Center, 
        Brookhaven National Laboratory, 
        Upton, NY 11973, USA}
\date{\today}
\pacs{12.38.Bx,12.15.Mm,12.38.Cy}
\begin{abstract}
Weak interactions between quarks give rise to hadronic parity violation which can be observed in nuclear and few-nucleon systems. 
We study the QCD renormalization of the isotensor component of parity violation at next-to-leading order accuracy. 
The renormalization group is employed to evolve the interaction down to hadronic scales.  
As the results are renormalization scheme dependent, 
we compare various schemes, 
including 't Hooft--Veltman dimensional regularization, and several regularization independent--momentum subtraction schemes. 
\end{abstract}
\maketitle

\section{Introduction}%

Measurement of parity violation in nuclear and few-nucleon reactions allows one to probe the flavor-conserving hadronic weak interaction. 
Experimentally hadronic parity violation has been observed in a variety of reactions,
see~\cite{Adelberger:1985ik,RamseyMusolf:2006dz} for reviews.
The interpretation of experimental data has been carried out using various models of the parity-violating nuclear force, 
the most popular of which is the meson-exchange model of Desplanques, Donoghue, and Holstein%
~\cite{Desplanques:1979hn}.
In this model, 
the parity-violating nuclear force is parametrized in terms of several parity-violating couplings between 
nucleons and mesons, 
which then give rise to parity violation in nuclei. 
Analyzed within this framework, 
constraints on parity-violating couplings coming from different experiments are not consistent. 
Such discrepancies could arise from several sources: 
uncertainty entering nuclear structure computations, 
model assumptions about the parity-violating nuclear force, 
dynamical effects due to the non-perturbative nature of QCD, 
\emph{etc}. 
Resolving this situation and connecting parity violation observed in nuclear reactions to parameters in the Standard Model is an ambitious undertaking.

Great progress has been made recently in describing parity-violating few-nucleon reactions in a model-independent fashion%
~\cite{Zhu:2004vw,Girlanda:2008ts,Phillips:2008hn,Shin:2009hi,Schindler:2009wd,Griesshammer:2011md,Vanasse:2011nd}. 
Ever advancing experiments in few-body systems, 
moreover, 
seek to better constrain the parity-violating nuclear force. 
Bounds have now been placed on hadronic parity violation in few-body reactions involving neutrons%
~\cite{Gericke:2011zz,Snow:2011zz}.
With these developments, 
there is cause to study parity violation theoretically using QCD. 
Lattice gauge theory computations can determine the parity-violating couplings between hadrons from first principles. 
Indeed, 
the first study of hadronic parity violation using lattice QCD has appeared%
~\cite{Wasem:2011tp}, 
wherein a technique to calculate the isovector parity-violating pion-nucleon coupling was explored.  
While lattice QCD calculations of parity violation are currently at quite an early stage, 
we are confident that precision information about hadronic parity violation will come from the lattice as refinements are made. 
To this end, 
one must understand the sources of parity violation at hadronic scales. 
The isovector channel has been studied at next-to-leading order%
~\cite{Tiburzi:2012hx}.
In this note, 
we focus on the next-to-leading order corrections in the isotensor channel.  
In this channel, 
parity violation gives rise to couplings between two pions and two nucleons, 
parity odd pion-photon-nucleon interactions%
~\cite{Kaplan:1992vj},
and an isotensor parity-violating nucleon-nucleon interaction.
From a computational perspective, 
this channel is expected to be statistically clean due to the lack of weak operator self-contractions.

We present our findings as follows. 
First in Sec.~\ref{s:First}, 
we discuss isotensor parity violation at the scale of weak interactions, 
and determine the four-quark interaction in this channel at next-to-leading order accuracy in the QCD coupling.  
The renormalization group is then employed in Sec.~\ref{s:Evo} to run the isotensor parity-violating 
interaction down to hadronic scales. 
As the evolution is renormalization scheme dependent, 
we present results for isotensor partiy violation using various renormalization schemes. 
We begin by obtaining results in 't Hooft-Veltman dimensional regularization. 
Results for several regularization independent--momentum subtraction schemes are determined and compared in
Sec.~\ref{s:summy}, 
which concludes this work.

\section{Isotensor Parity Violation} \label{s:First} %

In the Standard Model, 
isotensor parity violation arises from the exchange of 
$W$ 
and 
$Z$
bosons between quarks. 
At scales below the $W$ mass, 
$M_W$, 
we can integrate out the weak vector bosons to arrive at an effective theory of the form
\begin{equation} \label{eq:EFT}
\cL^{\D I = 2}_{PV} 
= 
\frac{G_F}{\sqrt{2}}
C(\mu) 
\cO (\mu)
.\end{equation}
There is only one isotensor parity violating four-quark operator~\cite{Dai:1991bx}, 
namely
\footnote{
To minimize statistical noise in lattice calculations, 
the chiral basis is not optimal because parity violation  arises  only upon taking the difference of hadronic matrix elements. 
A practical way to rewrite the isotensor parity-violating operator 
$\cO$ 
for use in lattice calculations is 
$\mathcal{O} 
= 
- 
\left[ 
(\ol q \, \tau^3 q)_A ( \ol q \, \tau^3 q)_V  
- 
\frac{1}{3} 
(\ol q \, \vec{\tau} q)_A \cdot 
( \ol q \, \vec{\tau} q)_V
\right]$,
where
$( \ol q q)_{A} (\ol q q)_V$
denotes the Lorentz contraction of axial-vector and vector bilinears,  
$( \ol q q)_{A} (\ol q q)_V = (\ol q \gamma_\mu \gamma_5 q ) ( \ol q \gamma^\mu q )$.
Written in this way, 
each term is parity violating. 
}
\begin{eqnarray}
\mathcal{O}
&=&
(\ol q \, \tau^3 \gamma_\mu q)_L ( \ol q \, \tau^3 \gamma^\mu q)_L
- 
\frac{1}{3} 
(\ol q \, \vec{\tau} \gamma_\mu q)_L \cdot ( \ol q \, \vec{\tau} \gamma^\mu q)_L
\notag \\
&& \phantom{placholdplachold}
- 
\Big\{ L \to R \Big\}
\label{c:O}
.\end{eqnarray}
The subscripts denote the left-handed and right-handed quark fields, 
$q_{L,R} = \frac{1}{2} (1 \mp \gamma_5) q$, 
with the isodoublet
$q = ( u , d)^T$.
In the absence of electromagnetism, 
the operator
$\cO$
does not mix with other parity violating operators. 
The corresponding Wilson coefficient,
$C(\mu)$, 
evolves with the scale parameter 
$\mu$
so that the effective theory described by
$\cL^{\D I = 2}_{PV}$
is scale independent.

%
%
\begin{figure}
\epsfig{file=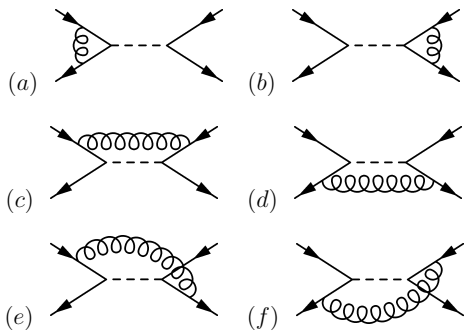,width=0.34\textwidth}
\caption{
One-loop QCD renormalization of weak-boson exchanges in the isotensor channel.
The straight lines are quarks, 
while the curly lines are gluons, 
and the dashed lines represent 
$W$ 
and 
$Z$ 
bosons. 
Wave-function renormalization contributes but is not depicted. 
}
\label{f:ZQCD}
\end{figure}
%
%

\begin{figure}
\epsfig{file=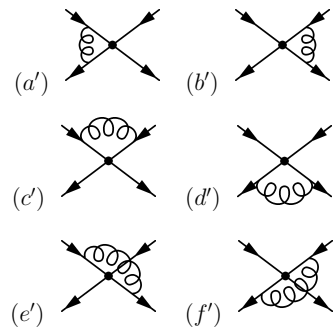,width=0.24\textwidth}
\caption{
One-loop QCD renormalization of the isotensor interaction in the parity-violating effective theory.
Filled circles denote the four-quark operator,
while all other diagram elements are as in Fig.~\ref{f:ZQCD}. 
Wave-function renormalization contributes but is not depicted. 
}
\label{f:HPVclasses}
\end{figure}

At tree-level, 
the Wilson coefficient is scale independent and receives contributions from both 
$W$ 
and 
$Z$
exchange. 
To separate these contributions, 
we write
$C^{(0)} = C^{(0)}_Z + C^{(0)}_W$, 
where the tree-level values of these coefficients are
$C^{(0)}_Z 
=
\frac{1}{2} - \sin^2 \theta_W$
and
$C^{(0)}_W
=
- \frac{1}{2} | V_{ud}|^2$.
At next-to-leading order, 
the Wilson coefficient of the operator 
$\cO$
is determined by comparing the one-loop renormalization in the full and effective theories. 
For a complete discussion of QCD renormalization beyond leading logarithms, 
see~\cite{Buchalla:1995vs}.
The gluon radiation in the full theory is depicted in 
Fig.~\ref{f:ZQCD}.
After the quark wave-function renormalization has been accounted for, 
the diagrams shown are finite. 
In order that the effective theory reproduces the full theory at one-loop, 
we must determine the matching coefficients by computing the effective theory diagrams shown in Fig.~\ref{f:HPVclasses}. 
These diagrams require regularization, 
and we begin by utilizing 't Hooft--Veltman dimensional regularization with 
$\overline{\text{MS}}$ 
subtraction. 
In this scheme, 
the axial-current vertex receives a finite renormalization at one loop~\cite{Buras:1989xd}, 
which indicates a non-vanishing two-loop anomalous dimension. 
To preserve chirality, 
we follow~\cite{Buras:1991jm,Ciuchini:1993vr}
and augment the 't Hooft--Veltman scheme with an additional multiplicative renormalization chosen 
to force the two-loop anomalous dimension of the axial-vector current to vanish.  
Additional renormalization schemes are considered in Sec.~\ref{s:summy}.

%
%
%
%

We choose to match the full and effective theories of isotensor parity violation at the scale
$\mu = M_W$. 
Diagrams with a 
$W$ 
boson exchanged are matched between theories without explicit logarithms in the matching coefficients. 
To one-loop order, 
we find
\begin{equation}
C_W (M_W) 
= 
C^{(0)}_W
\left[
1 - 
3 
\frac{\alpha_s (M_W)}{4 \pi}
\right]
.\end{equation}
On the other hand, 
diagrams in the full theory with a 
$Z$ 
boson exchanged will contribute logarithms of the form 
$\log \left(M_W / M_Z \right) = \log \left(\cos \theta_W \right)$
to the matching coefficients. 
Computing the matching coefficients for 
$Z$-exchange diagrams, 
we find
\begin{equation}
C_Z(M_W)
=
C^{(0)}_Z
\left[ 1 + \Bigg( 4 \log \left( \cos \theta_W \right) - 3 \Bigg) \frac{\alpha_s (M_W)}{4 \pi} \right]
.\end{equation}
Finally the desired initial value of the Wilson coefficient 
$C(\mu)$
at the 
$W$-boson
mass is simply
\begin{equation} \label{eq:initial}
C(M_W) 
= 
C_Z(M_W) + C_W(M_W)
,\end{equation}
with the individual terms specified above.

\section{Evolution} \label{s:Evo} %

The Wilson coefficient, 
$C(\mu)$, 
satisfies the renormalization group equation, 
$\mu \frac{d}{d\mu} C(\mu) 
= 
\Gamma(\mu) C(\mu)$,
where the anomalous dimension
$\Gamma(\mu)$
has the expansion to second order
$\Gamma = \gamma_0 \frac{\alpha_s}{4 \pi} + \gamma_1 \left( \frac{\alpha_s}{4 \pi} \right)^2
$,
with the scale dependence of 
$\Gamma$
arising from the running coupling, 
$\alpha_s(\mu)$. 
At next-to-leading order, 
the running of the QCD coupling is determined by the equation
$\mu^2 \frac{d}{d \mu^2} \frac{\alpha_s}{4\pi}
=
- \beta_0 
\left(
\frac{\alpha_s}{4 \pi}
\right)^2
- 
\beta_1 
\left(
\frac{\alpha_s}{4 \pi}
\right)^3
$,
with
$\beta_0 = 11 - \frac{2}{3} N_f$, 
and 
$\beta_1 = 102 -  \frac{38}{3}N_f$.

The anomalous dimension 
$\Gamma$
of the isotensor parity-violating operator,
$\cO$,
can be deduced from known results concerning the QCD renormalization of 
$|\Delta S| = 1$ 
operators.
Following~\cite{Tiburzi:2012hx}, 
we observe that parity invariance of QCD and of the renormalization scheme allows us to determine
the anomalous dimension of 
$\cO$
in Eq.~\eqref{c:O} 
by considering the
$V_L \otimes V_L$
combination of quark bilinears. 
Due to the isotensor nature of the operator, 
moreover, 
there are no penguin contractions to consider in the QCD renormalization. 
The anomalous dimensions arising from current-current contractions of 
$V_L \otimes V_L$
operators have been determined in~\cite{Buras:1992tc}
up to next-to-leading order. 
A closed basis of 
$V_L \otimes V_L$ 
operators with only current-current contractions is
\begin{eqnarray}
O_1
&=&
( \ol \psi_1 \gamma_\mu \psi_2)_L 
( \ol \psi_3 \gamma^\mu \psi_4)_L
\notag \\
O_2 
&=&
( \ol \psi_1 \gamma_\mu \psi_2 \, ]_L 
[ \, \ol \psi_3 \gamma^\mu \psi_4 )_L
\label{eq:ops}
,\end{eqnarray}
where the 
$\psi_i$ 
are each distinguishable quark fields, 
and the mixed brackets denote the color contraction
$( \ol \psi_1 \psi_2 \, ] \, [ \, \ol \psi_3 \psi_4 ) = \sum_{ab} \ol \psi {}^a_1 \psi^b_2 \, \ol \psi {}_3^b \psi_4^a$.
For the isotensor operator 
$\cO$, 
the analogous color rearranged operator
$\tilde{\cO}$
is identical to 
$\cO$
after a Fierz transformation. 
Because the 't Hooft--Veltman scheme respects Fierz transformations~\cite{Buras:1991jm}, 
mixing of 
$\cO$ 
into 
$\tilde{\cO}$ 
is simply equivalent to a renormalization of 
$\cO$. 
With the anomalous dimension matrix 
$\Gamma_{Q_{\text{VLL}}}$
of the two operators in Eq.~\eqref{eq:ops}, 
the desired anomalous dimension
$\Gamma$
of 
$\cO$
is merely the sum of the first row of 
$\Gamma_{Q_{\text{VLL}}}$. 
Hence we find
$\gamma_0 = 4$,
and
$\gamma_1 = \frac{419}{3} - \frac{76}{9} N_f$.

At next-to-leading order, 
the solution to the renormalization group equation is
$C(\mu^\prime)
= 
U(\mu^\prime, \mu) C(\mu)
$,
with 
\begin{eqnarray}
U(\mu', \mu)
&=&
\left[ 
\frac{\alpha_s(\mu^\prime)}{\alpha_s(\mu)}
\right]^{- \frac{\gamma_0}{2 \beta_0}}
\left[
1 
+
\frac{\alpha_s(\mu^\prime) - \alpha_s(\mu)}{4 \pi}
J
\right], \,\,\,
\label{eq:approximation}
\end{eqnarray}
where
$J =  \frac{\gamma_0 \beta_1}{2 \beta^2_0} - \frac{\gamma_1}{2 \beta_0}$.
Each of the quantities, 
$\beta_0$, 
$\beta_1$, 
and
$\gamma_1$,
depends on the number of active flavors, 
$N_f$. 
This number changes as we cross heavy quark thresholds. 
As a result, 
the Wilson coefficient at hadronic scales, 
$\Lambda_{\text{QCD}} < \mu < m_c$, 
is given by
\begin{equation} \label{eq:run}
C(\mu) 
= 
U^{(3)}(\mu, m_c)
U^{(4)}(m_c, m_b)
U^{(5)}(m_b, M_W)
C(M_W)
.\end{equation}
There are no one-loop matching conditions required as we cross heavy quark thresholds. 
To this order, 
it is sufficient to account for these thresholds via the scale parameters 
$\Lambda^{(N_f)}$
entering the definition of the two-loop running coupling, 
$\alpha_s^{(N_f)}(\mu, \Lambda^{(N_f)})$. 
The scale parameters are determined by enforcing continuity of the coupling 
at the heavy quark threshold, 
$m_Q$, 
namely
$\a_s^{(N_f)}(m_Q, \Lambda^{(N_f)})
=
\a_s^{(N_f -1)} (m_Q, \Lambda^{(N_f - 1)})
$.
Consequently 
the Wilson coefficient is described by a piece-wise continuous function, 
$C(\mu)$, 
of the renormalization scale
$\mu$.

\section{Summary of Results} \label{s:summy}%

Above we discuss parity violation in the isotensor channel at the scale of weak interactions, 
and the QCD renormalization group evolution to lower scales. 
Now we determine the strength of the parity-violating isotensor operator at hadronic scales 
in 't Hooft-Veltman dimensional regularization. 
Further schemes are discussed, 
and the corresponding results for isotensor parity violation are also determined in these schemes.

With the value of the Wilson coefficient at the weak scale in 't Hooft-Veltman regularization, 
i.e.~$C(M_W)$ in Eq.~\eqref{eq:initial}, 
we can run down to hadronic scales using the solution to the renormalization group equation,
Eq.~\eqref{eq:run}.
This result is presented in Table~\ref{t:results}, 
and requires the
$\ol{\text{MS}}$ 
masses of heavy quarks, 
and the value of 
$\alpha_s(M_Z)$, 
as described in~\cite{Tiburzi:2012hx}.
In this scheme, 
the next-to-leading order corrections are 
$\sim 20\%$
at the scale 
$\mu = 1 \, \texttt{GeV}$.
\footnote{
Our leading-order value is different than that obtained in~\cite{Kaplan:1992vj}, 
$C(1 \, \texttt{GeV}) /  C^{(0)} = 0.79$.
The difference arises from two sources. 
Firstly 
heavy quark masses were largely uncertain at that time. 
Secondly, 
we use the two-loop running of 
$\alpha_s$ 
in our leading-order result. 
Using historical values for input parameters~\cite{PDG:1992} along with their uncertainties, 
we obtain 
$C(1 \, \texttt{GeV}) /  C^{(0)} = 0.78(1)$
from one-loop running, 
consistent with~\cite{Kaplan:1992vj}.
Current-day uncertainties on input parameters lead to uncertainties on the Wilson coefficient that are an order of magnitude smaller.
} 

The conversion to other renormalization schemes can be achieved through one-loop matching. 
For example, 
the value of 
$C_X(\mu)$
in scheme 
$X$
is found from that calculated in the 't Hooft-Veltman scheme, 
$C_{HV}(\mu)$, 
via the relation
$C_X(\mu) = \left[ 1 + \Delta r \frac{\alpha_s(\mu)}{4 \pi} \right] C_{HV}(\mu)$, 
where 
$\Delta r$
is the difference of finite contributions to one-loop diagrams between schemes, 
$\Delta r = r_{X} - r_{HV}$. 
For the isotensor parity violating operator, 
these differences are known for all renormalization schemes we consider. 
It is common to quote results in na\"ive dimensional regularization, 
in which 
$\gamma_5$ 
remains anti-commuting in 
$d$-dimensions. 
At two-loop order, 
consistent results for the renormalization of four-quark operators can be obtained in this scheme, 
see, 
for example,~\cite{Buras:1992tc}. 
From one-loop matching, 
the Wilson coefficient of isotensor parity violation using na\"ive dimensional regularization is determined in 
Table~\ref{t:results}, 
and differs by 
$\sim 20\%$
from that in 't Hooft-Veltman dimensional regularization. 
This is due to a relatively large matching coefficient compounded with the low scale 
$\mu = 1 \, \texttt{GeV}$. 
The next-to-leading correction in na\"ive dimensional regularization is 
$\sim 5\%$ 
of the leading-order value.

\begin{table}
\caption{\label{t:results}
Wilson coefficient of the 
$| \Delta I | = 2$ 
parity-violating operator, 
$C (\mu)$, 
at hadronic scales.
We compare the leading-order 
value 
with next-to-leading results obtained in various renormalization schemes. 
In each case, 
we scale the coefficient by its tree-level value, 
$C^{(0)}$. 
All momentum subtraction schemes employ Landau gauge. 
}
\begin{center}
\begin{tabular}{|c|c|}
\hline 
\hline
& 
$\qquad C(1 \, \texttt{GeV}) /  C^{(0)} \qquad$ 
\tabularnewline
\hline
\hline
Leading Order
&
$0.70$
\tabularnewline
\hline
't Hooft-Veltman
&
$0.58$
\tabularnewline
Na\"ive Dim.~Reg.
&
$0.74$
\tabularnewline
RI/MOM
&
$0.77$
\tabularnewline
RI/SMOM$(\gamma_\mu, \rlap \slash q)$
&
$0.67$
\tabularnewline
RI/SMOM$(\gamma_\mu, \gamma_\mu)$
&
$0.75$
\tabularnewline
RI/SMOM$(\rlap \slash q, \rlap \slash q)$
&
$0.73$
\tabularnewline
RI/SMOM$(\rlap \slash q, \gamma_\mu)$
&
$0.81$
\tabularnewline
\hline
\hline
\end{tabular} 
\end{center} 
\end{table}

While dimensional regularization schemes are ideal for perturbative calculations, 
schemes that can be defined non-perturbatively are efficacious for lattice calculations, 
such as the regularization independent--momentum scheme
(RI/MOM)~\cite{Martinelli:1994ty}. 
In this scheme, 
each external leg of the four-quark operator is taken with an identical off-shell momentum. 
Results in the RI/MOM scheme can similarly be determined from one-loop matching, 
and the resulting Wilson coefficient is given in the table. 
The choice of quark momenta in this scheme is quite exceptional, 
for example, no momentum flows out of the operator. 
It has been argued that an exceptional renormalization point is undesirable from chiral symmetry breaking considerations, 
and renormalization about an exceptional point generally exhibits poor convergence%
~\cite{Aoki:2007xm,Sturm:2009kb}.
To correct these maladies, 
schemes with 
a symmetric momentum subtraction point
(RI/SMOM) 
have been suggested in which momentum flows out of the operator.
Results for the four different RI/SMOM schemes proposed for four-quark operators in%
~\cite{Aoki:2010pe} 
(see also~\cite{Lehner:2011fz})
are given in Table~\ref{t:results}.

The next-to-leading order results we present are essential  
to study isotensor parity violation in QCD. 
As with the investigation of other weak interaction processes using the lattice method, 
we are limited by the convergence of perturbation theory. 
While non-perturbative schemes can be used to compute operator matrix elements, 
the scale of such lattice computations is limited to a few 
$\texttt{GeV}$. 
As we require input from perturbative calculations to determine the interaction strength, 
the scale independence of the effective theory, 
Eq.~\eqref{eq:EFT}, 
is hence limited by the efficacy of perturbation theory. 
We compare several schemes above, 
and find on average
$\sim 10\%$
differences arise, 
which gives a rough indication as to the size of higher-order perturbative corrections.

Our study of isotensor parity violation shows that QCD evolution 
generally suppresses the magnitude of the interaction at hadronic scales about
$25 \%$. 
It will be interesting to see if hadronic matrix elements are small in this channel compared to the 
$\Delta I = 0$, 
$1$ 
channels.
If this were the case, 
hadronic parity violation would parallel the 
$\Delta I = \frac{1}{2}$ rule, 
for which the maximal isospin changing channel is considerably suppressed. 
For this comparison, 
we must ultimately await the evaluation of non-perturbative physics from lattice QCD.

\begin{acknowledgments}
Work supported in part by a joint CCNY--RBRC fellowship, 
and the U.S.~National Science Foundation, 
under Grant No.~PHY-$1205778$.
\end{acknowledgments}

\appendix

\bibliography{hb}

\end{document}